\documentstyle[12pt,graphicx]{article}

\def\mathfrak{\bf}

\def\be{\begin{equation}}
\def\ee{\end{equation}}
\def\bea{\begin{eqnarray}}
\def\eea{\end{eqnarray}}

\def\dt#1{\on{\hbox{\bf .}}{#1}}                
\def\Dot#1{\dt{#1}}
\def\IR{\relax{\rm I\kern-.18em R}}
\def\binomial#1#2{\left(\,{\buildrel
{\raise4pt\hbox{$\displaystyle{#1}$}}\over
{\raise-6pt\hbox{$\displaystyle{#2}$}}}\,\right)}

\def\[{\lfloor{\hskip 0.35pt}\!\!\!\lceil}
\def\]{\rfloor{\hskip 0.35pt}\!\!\!\rceil}

\newcommand{\AmS}{{\protect\the\textfont2
  A\kern-.1667em\lower.5ex\hbox{M}\kern-.125emS}}

\catcode`@=11
\def\un#1{\relax\ifmmode\@@underline#1\else
        $\@@underline{\hbox{#1}}$\relax\fi}
\catcode`@=12

\def\fracm#1#2{\hbox{\large{${\frac{{#1}}{{#2}}}$}}}

\def\ad{{\kern0.5pt
                   \alpha \kern-5.05pt
\raise5.8pt\hbox{$\textstyle.$}\kern
0.5pt}}

\def\Dot#1{{\kern0.5pt
     {#1} \kern-5.05pt \raise5.8pt\hbox{$\textstyle.$}\kern
0.5pt}}

\def\a{\alpha}
\def\b{\beta}

\def\d{\delta}
\def\e{\epsilon}

\def\g{\gamma}

\def\i{\iota}

\def\q{\theta}

\def\s{\sigma}

\def\S{\Sigma}

\def\cf{{\cal F}}

\def\cm{{\cal M}}

\def\car{{\cal R}}

\def\cy{{\cal Y}}

\def\bo{{\raise.15ex\hbox{\large$\Box$}}}               
\def\de{\nabla}                                         
\def\TH{{\raise.2ex\hbox{$\displaystyle \bigodot$}\mskip-4.7mu \llap H
\;}}
\def\face{{\raise.2ex\hbox{$\displaystyle \bigodot$}\mskip-2.2mu \llap
{$\ddot
        \smile$}}}                                      

\def\sp#1{{}^{#1}}                              
\def\sb#1{{}_{#1}}                              
\def\Bar#1{\overline{#1}}                       
\def\leftrightarrowfill{$\mathsurround=0pt \mathord\leftarrow \mkern-6mu
        \cleaders\hbox{$\mkern-2mu \mathord- \mkern-2mu$}\hfill
        \mkern-6mu \mathord\rightarrow$}
\def\dvec#1{\vbox{\ialign{##\crcr
        \leftrightarrowfill\crcr\noalign{\kern-1pt\nointerlineskip}
        $\hfil\displaystyle{#1}\hfil$\crcr}}}           
\def\dt#1{{\buildrel {\hbox{\LARGE .}} \over {#1}}}     

\def\fracm#1#2{\hbox{\large{${\frac{{#1}}{{#2}}}$}}}
\def\frac#1#2{{\textstyle{#1\over\vphantom2\smash{\raise.20ex
        \hbox{$\scriptstyle{#2}$}}}}}                   
\def\ha{\frac12}                                        
\def\sfrac#1#2{{\vphantom1\smash{\lower.5ex\hbox{\small$#1$}}\over
        \vphantom1\smash{\raise.4ex\hbox{\small$#2$}}}} 
\def\bfrac#1#2{{\vphantom1\smash{\lower.5ex\hbox{$#1$}}\over
        \vphantom1\smash{\raise.3ex\hbox{$#2$}}}}       
\def\afrac#1#2{{\vphantom1\smash{\lower.5ex\hbox{$#1$}}\over#2}}    
\def\on#1#2{\mathop{\null#2}\limits^{#1}}               

\newskip\humongous \humongous=0pt plus 1000pt minus 1000pt
\def\caja{\mathsurround=0pt}
\def\eqalign#1{\,\vcenter{\openup2\jot \caja
        \ialign{\strut \hfil$\displaystyle{##}$&$
        \displaystyle{{}##}$\hfil\crcr#1\crcr}}\,}
\newif\ifdtup

  \def\pp{{\mathchoice
          {
              \kern 1pt%
              \raise 1pt
              \vbox{\hrule width5pt height0.4pt depth0pt
                    \kern -2pt
                    \hbox{\kern 2.3pt
                          \vrule width0.4pt height6pt depth0pt
                          }
                    \kern -2pt
                    \hrule width5pt height0.4pt depth0pt}%
                    \kern 1pt
           }
            {
              \kern 1pt%
              \raise 1pt
              \vbox{\hrule width4.3pt height0.4pt depth0pt
                    \kern -1.8pt
                    \hbox{\kern 1.95pt
                          \vrule width0.4pt height5.4pt depth0pt
                          }
                    \kern -1.8pt
                    \hrule width4.3pt height0.4pt depth0pt}%
                    \kern 1pt
            }
            {
              \kern 0.5pt%
              \raise 1pt
              \vbox{\hrule width4.0pt height0.3pt depth0pt
                    \kern -1.9pt  
                    \hbox{\kern 1.85pt
                          \vrule width0.3pt height5.7pt depth0pt
                          }
                    \kern -1.9pt
                    \hrule width4.0pt height0.3pt depth0pt}%
                    \kern 0.5pt
            }
            {
              \kern 0.5pt%
              \raise 1pt
              \vbox{\hrule width3.6pt height0.3pt depth0pt
                    \kern -1.5pt
                    \hbox{\kern 1.65pt
                          \vrule width0.3pt height4.5pt depth0pt
                          }
                    \kern -1.5pt
                    \hrule width3.6pt height0.3pt depth0pt}%
                    \kern 0.5pt
            }
        }}

  \def\mm{{\mathchoice
                       {
                             \kern 1pt
               \raise 1pt    \vbox{\hrule width5pt height0.4pt depth0pt
                                  \kern 2pt
                                  \hrule width5pt height0.4pt depth0pt}
                             \kern 1pt}
                       {
                            \kern 1pt
               \raise 1pt \vbox{\hrule width4.3pt height0.4pt depth0pt
                                  \kern 1.8pt
                                  \hrule width4.3pt height0.4pt depth0pt}
                             \kern 1pt}
                       {
                            \kern 0.5pt
               \raise 1pt
                            \vbox{\hrule width4.0pt height0.3pt depth0pt
                                  \kern 1.9pt
                                  \hrule width4.0pt height0.3pt depth0pt}
                            \kern 1pt}
                       {
                           \kern 0.5pt
             \raise 1pt  \vbox{\hrule width3.6pt height0.3pt depth0pt
                                  \kern 1.5pt
                                  \hrule width3.6pt height0.3pt depth0pt}
                           \kern 0.5pt}
                       }}

\def\pd{{\kern0.5pt
                   + \kern-5.05pt \raise5.8pt\hbox{$\textstyle.$}\kern
0.5pt}}

\def\pmd{{\kern0.5pt
                  \pm \kern-5.05pt \raise6.3pt\hbox{$\textstyle.$}\kern1.5pt}}

\def\md{{\mathchoice
   {
      {{\kern 1pt - \kern-6.2pt \raise5pt\hbox{$\textstyle.$}\kern 1pt}}}
    {
      {{\kern 1pt - \kern-6.2pt \raise5pt\hbox{$\textstyle.$}\kern 1pt}}}
    {
      {\kern0.5pt - \kern-5.05pt \raise3.4pt\hbox{$\textstyle.$}\kern0.5pt}}
    {
      {\kern0.5pt - \kern-5.05pt \raise3.4pt\hbox{$\textstyle.$}\kern0.5pt}}}}

\def\ad{{\dot{\alpha}}}

\def\pp{{\mathchoice
          {
              \kern 1pt%
              \raise 1pt
              \vbox{\hrule width5pt height0.4pt depth0pt
                    \kern -2pt
                    \hbox{\kern 2.3pt
                          \vrule width0.4pt height6pt depth0pt
                          }
                    \kern -2pt
                    \hrule width5pt height0.4pt depth0pt}%
                    \kern 1pt
           }
            {
              \kern 1pt%
              \raise 1pt
              \vbox{\hrule width4.3pt height0.4pt depth0pt
                    \kern -1.8pt
                    \hbox{\kern 1.95pt
                          \vrule width0.4pt height5.4pt depth0pt
                          }
                    \kern -1.8pt
                    \hrule width4.3pt height0.4pt depth0pt}%
                    \kern 1pt
            }
            {
              \kern 0.5pt%
              \raise 1pt
              \vbox{\hrule width4.0pt height0.3pt depth0pt
                    \kern -1.9pt  
                    \hbox{\kern 1.85pt
                          \vrule width0.3pt height5.7pt depth0pt
                          }
                    \kern -1.9pt
                    \hrule width4.0pt height0.3pt depth0pt}%
                    \kern 0.5pt
            }
            {
              \kern 0.5pt%
              \raise 1pt
              \vbox{\hrule width3.6pt height0.3pt depth0pt
                    \kern -1.5pt
                    \hbox{\kern 1.65pt
                          \vrule width0.3pt height4.5pt depth0pt
                          }
                    \kern -1.5pt
                    \hrule width3.6pt height0.3pt depth0pt}%
                    \kern 0.5pt
            }
        }}

  \def\mm{{\mathchoice
                       {
                             \kern 1pt
               \raise 1pt    \vbox{\hrule width5pt height0.4pt depth0pt
                                  \kern 2pt
                                  \hrule width5pt height0.4pt depth0pt}
                             \kern 1pt}
                       {
                            \kern 1pt
               \raise 1pt \vbox{\hrule width4.3pt height0.4pt depth0pt
                                  \kern 1.8pt
                                  \hrule width4.3pt height0.4pt depth0pt}
                             \kern 1pt}
                       {
                            \kern 0.5pt
               \raise 1pt
                            \vbox{\hrule width4.0pt height0.3pt depth0pt
                                  \kern 1.9pt
                                  \hrule width4.0pt height0.3pt depth0pt}
                            \kern 1pt}
                       {
                           \kern 0.5pt
             \raise 1pt  \vbox{\hrule width3.6pt height0.3pt depth0pt
                                  \kern 1.5pt
                                  \hrule width3.6pt height0.3pt depth0pt}
                           \kern 0.5pt}
                       }}

\def\pd{{\kern0.5pt
                   + \kern-5.05pt \raise5.8pt\hbox{$\textstyle.$}\kern
0.5pt}}

\def\pmd{{\kern0.5pt
                  \pm \kern-5.05pt \raise6.3pt\hbox{$\textstyle.$}\kern1.5pt}}

\def\md{{\mathchoice
   {
      {{\kern 1pt - \kern-6.2pt \raise5pt\hbox{$\textstyle.$}\kern 1pt}}}
    {
      {{\kern 1pt - \kern-6.2pt \raise5pt\hbox{$\textstyle.$}\kern 1pt}}}
    {
      {\kern0.5pt - \kern-5.05pt \raise3.4pt\hbox{$\textstyle.$}\kern0.5pt}}
    {
      {\kern0.5pt - \kern-5.05pt \raise3.4pt\hbox{$\textstyle.$}\kern0.5pt}}}}

\def\dslash{\not{\hbox{\kern-2pt $\partial$}}}
\def\Dslash{\not{\hbox{\kern-4pt $D$}}}
\def\pslash{\not{\hbox{\kern-2.3pt $p$}}}
 \newtoks\slashfraction
 \slashfraction={.13}
 \def\slash#1{\setbox0\hbox{$ #1 $}
 \setbox0\hbox to \the\slashfraction\wd0{\hss \box0}/\box0 }

\font\ro=cmsy10                          
\def\kcr{{\hbox{\ro \char'170}}}                
\def\ktl{{\hbox{\ro \char'170}}}        
\def\ktr{{\hbox{\ro \char'170}}}        
\def\kbl{{\hbox{\ro \char'170}}}        
\def\kbr{{\hbox{\ro \char'170}}}        

\def\plpl{\raise-2pt\hbox{$\raise3pt\hbox{$_+$}\hskip-6.67pt\raise0.0pt
\hbox{$^+$}\hskip 0.01pt$}}
\def\mimi{\raise-2pt\hbox{$\raise3pt\hbox{$_-$}\hskip-6.67pt\raise0.0pt
\hbox{$^-$}\hskip 0.01pt$}}

\def\bo{{\raise.15ex\hbox{\large$\Box$}}}               
\def\de{\nabla}                                         
\def\TH{{\raise.2ex\hbox{$\displaystyle \bigodot$}\mskip-4.7mu \llap H \;}}
\def\face{{\raise.2ex\hbox{$\displaystyle \bigodot$}\mskip-2.2mu \llap {$\ddot
        \smile$}}}                                      

\def\sp#1{{}^{#1}}                              
\def\sb#1{{}_{#1}}                              
\def\Bar#1{\overline{#1}}                       
\def\leftrightarrowfill{$\mathsurround=0pt \mathord\leftarrow \mkern-6mu
        \cleaders\hbox{$\mkern-2mu \mathord- \mkern-2mu$}\hfill
        \mkern-6mu \mathord\rightarrow$}
\def\dvec#1{\vbox{\ialign{##\crcr
        \leftrightarrowfill\crcr\noalign{\kern-1pt\nointerlineskip}
        $\hfil\displaystyle{#1}\hfil$\crcr}}}           
\def\dt#1{{\buildrel {\hbox{\LARGE .}} \over {#1}}}     

\def\fracm#1#2{\hbox{\large{${\frac{{#1}}{{#2}}}$}}}
\def\frac#1#2{{\textstyle{#1\over\vphantom2\smash{\raise.20ex
        \hbox{$\scriptstyle{#2}$}}}}}                   
\def\ha{\frac12}                                        
\def\sfrac#1#2{{\vphantom1\smash{\lower.5ex\hbox{\small$#1$}}\over
        \vphantom1\smash{\raise.4ex\hbox{\small$#2$}}}} 
\def\bfrac#1#2{{\vphantom1\smash{\lower.5ex\hbox{$#1$}}\over
        \vphantom1\smash{\raise.3ex\hbox{$#2$}}}}       
\def\afrac#1#2{{\vphantom1\smash{\lower.5ex\hbox{$#1$}}\over#2}}    
\def\on#1#2{\mathop{\null#2}\limits^{#1}}               

\parskip=\medskipamount                 
\thicklines                         

\def\oldheadpic{                                
        \setlength{\unitlength}{.4mm}
        \thinlines
        \par
        \begin{picture}(349,16)
        \put(325,16){\line(1,0){4}}
        \put(330,16){\line(1,0){4}}
        \put(340,16){\line(1,0){4}}
        \put(335,0){\line(1,0){4}}
        \put(340,0){\line(1,0){4}}
        \put(345,0){\line(1,0){4}}
        \put(329,0){\line(0,1){16}}
        \put(330,0){\line(0,1){16}}
        \put(339,0){\line(0,1){16}}
        \put(340,0){\line(0,1){16}}
        \put(344,0){\line(0,1){16}}
        \put(345,0){\line(0,1){16}}
        \put(329,16){\oval(8,32)[bl]}
        \put(330,16){\oval(8,32)[br]}
        \put(339,0){\oval(8,32)[tl]}
        \put(345,0){\oval(8,32)[tr]}
        \end{picture}
        \par
        \thicklines
        \vskip.2in}
\def\oldtitle#1#2#3#4{\oldheadpic\begin{center}\vglue.5in{\large\bf #1}\\[.6in]
        {#2}\\[.1in] {\it Department of Physics and Astronomy}\\
        {\it University of Maryland, College Park, MD 20742}\\[.6in]
        Physics Publication \#{#3}\\ {#4}\\[1.5in] {\bf ABSTRACT}\\[.1in]
        \end{center} \begin{quotation}}                 
\def\oldTitle#1#2#3#4#5#6#7{\oldheadpic\begin{center} \vglue .4in
        {\large\bf #1}\\[.4in]
        {#2}\\[.1in] {\it Department of Physics and Astronomy}\\
        {\it University of Maryland, College Park, MD 20742}\\[.1in]
        {#3}\\[.1in] {\it {#4}}\\ {\it {#5}}\\[.4in]
        Physics Publication \#{#6}\\ {#7}\\[.5in] {\bf ABSTRACT}\\[.1in]
        \end{center} \begin{quotation}}                 
\def\border{                                            
        \setlength{\unitlength}{1mm}
        \newcount\xco
        \newcount\yco
        \xco=-21
        \yco=12
        \begin{picture}(140,0)
        \put(\xco,\yco){$\ktl$}
        \advance\yco by-1
        {\loop
        \put(\xco,\yco){$\kcr$}
        \advance\yco by-2
        \ifnum\yco>-240
        \repeat
        \put(\xco,\yco){$\kbl$}}
        \xco=158
        \yco=12
        \put(\xco,\yco){$\ktr$}
        \advance\yco by-1
        {\loop
        \put(\xco,\yco){$\kcr$}
        \advance\yco by-2
        \ifnum\yco>-240
        \repeat
        \put(\xco,\yco){$\kbr$}}
        \put(-20,13){\tiny **University of Maryland * Center for String and
         Particle  Theory* Physics Department***University of Maryland *Center
        for String and Particle  Theory** }
        \put(-20,-241.5){\tiny **University of Maryland * Center for String and
         Particle  Theory* Physics Department***University of Maryland *Center
        for String and Particle  Theory** }
        \end{picture}
        \par\vskip-8mm}
\def\bordero{                                           
        \setlength{\unitlength}{1mm}
        \newcount\xco
        \newcount\yco
        \xco=-31
        \yco=12
        \begin{picture}(140,0)
        \put(\xco,\yco){$\ktl$}
        \advance\yco by-1
        {\loop
        \put(\xco,\yco){$\kclr}
        \advance\yco by-2
        \ifnum\yco>-240
        \repeat
        \put(\xco,\yco){$\kbl$}}
        \xco=151
        \yco=12
        \put(\xco,\yco){$\ktr$}
        \advance\yco by-1
        {\loop
        \put(\xco,\yco){$\kcr$}
        \advance\yco by-2
        \ifnum\yco>-240
        \repeat
        \put(\xco,\yco){$\kbr$}}
        \put(-20,12){\ooo bacdefghidfghghdhededbihdgdfdfhhdheidhdhebaaahjhhdahba

hgdedge
   hgfdiehhgdigicba}
        \put(-20,-241.5){\ooo ababaighefdbfghgeahgdfgafagihdidihiidhiagfedhadbfd

ecdcdfa
   gdcbhaddhbgfchbgfdacfediacbabab}
        \end{picture}
        \par\vskip-8mm}
\def\headpic{                                           
        \indent
        \setlength{\unitlength}{.4mm}
        \thinlines
        \par
        \begin{picture}(29,16)
        \put(165,16){\line(1,0){4}}
        \put(170,16){\line(1,0){4}}
        \put(180,16){\line(1,0){4}}
        \put(175,0){\line(1,0){4}}
        \put(180,0){\line(1,0){4}}
        \put(185,0){\line(1,0){4}}
        \put(169,0){\line(0,1){16}}
        \put(170,0){\line(0,1){16}}
        \put(179,0){\line(0,1){16}}
        \put(180,0){\line(0,1){16}}
        \put(184,0){\line(0,1){16}}
        \put(185,0){\line(0,1){16}}
        \put(169,16){\oval(8,32)[bl]}
        \put(170,16){\oval(8,32)[br]}
        \put(179,0){\oval(8,32)[tl]}
        \put(185,0){\oval(8,32)[tr]}
        \end{picture}
        \par\vskip-6.5mm
        \thicklines}
\def\title#1#2#3#4{\border\headpic {\hbox to\hsize{#4 \hfill UMDEPP #3}}\par
        \begin{center} \vglue .5in {\large\bf #1}\\[.6in]
        {#2}\\[.1in] {\it Department of Physics and Astronomy}\\
        {\it University of Maryland, College Park, MD 20742}\\[1.5in]
        {\bf ABSTRACT}\\[.1in] \end{center} \begin{quotation}}  
\def\Title#1#2#3#4#5#6#7{\border\headpic
        {\hbox to\hsize{#7 \hfill UMDEPP #6}}\par
        \begin{center} \vglue .4in {\large\bf #1}\\[.4in]
        {#2}\\[.1in] {\it Department of Physics and Astronomy}\\
        {\it University of Maryland, College Park, MD 20742}\\[.1in]
        {#3}\\[.1in] {\it {#4}}\\ {\it {#5}}\\[.5in] {\bf ABSTRACT}\\[.1in]
        \end{center} \begin{quotation}}                 
\def\endtitle{\end{quotation}\newpage}                  

\def\qd{{\kern0.5pt
                   q \kern-5.05pt \raise5.8pt\hbox{$\textstyle.$}\kern
0.5pt}}

\begin{document}

\def\dt#1{\on{\hbox{\bf .}}{#1}}                
\def\Dot#1{\dt{#1}}

\def\gfrac#1#2{\frac {\scriptstyle{#1}}
        {\mbox{\raisebox{-.6ex}{$\scriptstyle{#2}$}}}}
\def\gg{{\hbox{\sc g}}}
\border\headpic {\hbox to\hsize{January 2009 \hfill
{UMDEPP 08-024}}}
\par
{$~$ \hfill
{hep-th/0901.4165}}
\par

\setlength{\oddsidemargin}{0.3in}
\setlength{\evensidemargin}{-0.3in}
\begin{center}
\vglue .10in
{\large\bf A Derivation of an Off-Shell \\
$2$D, ${\cal N}=(2,2)$ Supergravity
\\[.1in]
Chiral Projection Operator\footnote
{Supported in part  by National Science Foundation Grant
PHY-0354401.}\  }
\\[.5in]

S.\, James Gates, Jr.\footnote{gatess@wam.umd.edu}
and Akin Morrison
\\[0.2in]

{\it Center for String and Particle Theory\\
Department of Physics, University of Maryland\\
College Park, MD 20742-4111 USA}\\[1.5in]

{\bf ABSTRACT}\\[.01in]
\end{center}
\begin{quotation}
{Utilizing the known off-shell formulation of 2D, $\cal N$ = (2,2) supergravity containing 
a finite number of auxiliary fields, there is shown to exist a simple form for a `chiral
projection operator' and an explicit expression for it is given.}

${~~~}$ \newline

\endtitle

\section{Introduction}

~~~~ Some years ago, the first discussion of a 2D, $\cal N$ = (2,2) supergravity theory
was given \cite{2DN4SGa} and elaborated upon in a subsequent work \cite{2DN4SGb}.
In both of these works, no discussion of the auxiliary fields required to close the
local supersymmetry algebra without resorting a set of equations of motion was
undertaken. Finally, thirteen years after the first such discussion, a formulation of
2D, $\cal N$ = 4 supergravity theory that included a {\em {finite}} number of auxiliary fields 
was presented \cite{2DN4SGc}.  The relation between these on-shell versus off-shell 
supergravity theories is not as direct as one might imagine.  The reason for this is the
 existence of a great plethora of 2D, $\cal N$ = 4 scalar multiplets \cite{2DN4SM}.
 Previous experience has shown that when this situation exists, the diverse scalar
 multiplets foreshadow distinct formulations of correspondingly diverse off-shell
 supergravity theories.
 
The existence of a relatively simple off-shell formulation of a 2D, $\cal N$ = 4 
supergravity theory implies that there should exist a straightforward way to completely
develop an {\em {efficient}} local integration theory for the associated local Salam-Strathdee 
superspace.  In this work, we will take the first major step in this direction by providing
the initial discussion of a local 2D, $\cal N$ = 4 chiral projection operator.

\section{A Review of an Off-Shell 2D, ${\cal N}=(2,2)$ Superspace
Supergravity Geometry}

~~~~ Let us begin by reviewing the results in \cite{2DN4SGc}. This work showed there exists 
component fields $ (e_a{}^m ,~ \psi_a {}^{\a i}, ~ A_{ai} {}^j , ~ B,  ~ G, ~H ) $ which describe
an off-shell 2D, ${\cal N}=(2,2)$ supergravity theory.  These are the components of that remain 
after imposing the following constraints on the 2D, $N$ = 4 superspace supergravity covariant 
derivative, (with $ \phi_{\a \, \b} \equiv - i [  C_{\a \b} G + i (\g^3
)_{\a \b} H ] $)
\begin{equation}
\begin{array}{lll}
~[\de \sb{\a i} , \de \sb{\b j} \} & = & 2 \Bar B [ \, C \sb {\a \b}
C \sb{ ij} \cm ~-~ ( \g \sp 3 ) \sb{ \a \b} \cy \sb{i j}] \,  ~~,  \\ 
~[\de \sb{\a i} , \Bar \de \sb{\b} \sp j \}   &=& 2  [~i  \d \sb i \sp j 
(\g \sp{c}) \sb{\a \b} \de \sb{c} ~+~  \d \sb i \sp j \phi \sb{\a}{}^{\g}
(\g^3)_{\g \b}  \cm  ~-~ i \phi \sb{\a \b} \cy \sb i \sp j  ~] ~~,   \\
~[\de \sb{\a i} , \de \sb{b} \}   &=& i {1 \over 2} \phi \sb {\a} \sp{\g}
(\g \sb b) \sb \g \sp {\b} \de \sb{\b i} ~+~ i {1 \over 2} (\g \sp 3 \g \sb b) 
\sb {\a} \sp{ \b} \Bar B C \sb{i j} \Bar \de \sb \b \sp j   \\     
&~&~~ -~ i(\g \sp 3 \g \sb b) \sb {\a \b}
\Bar \S \sp \b \sb i \cm ~+~ i (\g \sb b) \sb{\a \b} \Bar \S \sp \b \sb j
\cy \sb i \sp j ~~,  \\  
~[\de \sb{a} , \de \sb{b} \}   &=& - \ha \e \sb{a b} 
[ (\g \sp 3) \sb \a \sp \b \S \sp{ \a i} \de \sb{ \b i}
~+~ (\g \sp 3) \sb \a \sp \b \Bar \S \sp \a \sb i \Bar \de \sb \b \sp i
~+~ \car \cm ~+~ i \cf \sb i \sp j \cy \sb j \sp i ]~~ .   
\end{array}       \label{CommAlg}
\end{equation}
The consistency of the Bianchi identities constructed from the commutator
algebra above required the conditions, 
\begin{equation}
\begin{array}{lll}
\Bar \de \sb \a \sp i B &=&0~~~~~~,~~~
\de \sb{\a i} B = -2 C \sb{i j} ( \g \sp 3 ) \sb{ \a \b} \S \sp{ \b j}~~,
 \\
\de \sb{\a i} G &=& \Bar \S \sb{ \a i}~~~,~~~  
\de \sb{\a i} H = i( \g \sp 3 ) \sb{ \a} \sp{ \b} \Bar \S \sb{ \b i},
~~~, \\
\Bar \de \sb \a \sp i \S \sp{\b j} &=& i  C \sp{i j}
(\g \sp 3 \g \sp a) \sb {\a } \sp{ \b} \de \sb a B  ~~,  \\
\de \sb{\a i} \S \sp{\b j} &=& {1 \over 2} \d \sb \a \sp \b
\d \sb i \sp j [ \car ~-~ 2 G \sp 2 ~-~ 2 H \sp 2 ~-~ 2 B \Bar B ]
~+~ i (\g \sp 3) \sb \a \sp \b \cf \sb i \sp j  \\
& &+~ i {1 \over 2} \d \sb i \sp j (\g \sp a ) \sb \a \sp \b (\de \sb a G) 
-~  {1 \over 2} \d \sb i \sp j (\g \sp 3 \g \sp a ) \sb \a \sp \b 
(\de \sb a H) ~~~.     
\end{array}     \label{BIs}
\end{equation}
The component gauge fields occur in the above supertensors in the 
following manner.  
\begin{equation}
\begin{array}{lll} 
\car {\big |} &=& \e^{a b} {\large \{ } ~ {\car}_{a b} (\hat \omega) ~+~  
[ ~ i 2 (\g^3 \g_a) _{\a \b} \psi_b {}^{\a i} \Bar \S^\b {}_i ~+~ {\rm 
{h.c.}} ~  ]  \\ 
& &~~~~~~ +~ 4 \phi_\a {}^\g (\g^3)_{\g \b} \psi_a {}^{\a i} {\Bar \psi}_b 
{}^\b {}_i  ~-~2  [~ C_{i j} \Bar B \psi_a {}^{\a i}  \psi_{b \a}{}^{j}
~+~ {\rm {h.c.}} ~  ]~ {\large \} } ~~, \\
&   & \\
\S ^{\a i} {\big |} & = & \e^{ab} {\large \{ } ~ \psi_{ab} {}^{\b i}
(\g^3)_\b {}^\a ~-~ i \psi_a {}^{\b i} {\phi}_\b {}^\g 
( \g^3 \g_b)_\g {}^\a   ~+~ i C^{ij} B \Bar \psi_a {}^\b {}_j 
(\g_b)_\b {}^\a  ~ {\large \} } ~~, \\ 
&   & \\
\cf_i {}^j {\big |} &=& \e^{ab} {\large \{ } ~ {\rm F}_{a b}(A)_{i}{}^j 
~-~ i 2  (\g_a)_{\a \b}  [~ \psi_b {}^{\a j} \Bar \S^\b {}_i ~+~ \Bar \psi_b 
{}^\a {}_i \S^{\b j} \\ 
& &~~~~~~~~~~~~~~~~~~-~ \frac 12 \d_i^j (\psi_b {}^{\a k} \Bar \S^\b {}_k 
~+~  \Bar \psi _b {}^\a {}_k \S^{\b k}) ~] \\
& & ~~~~~~~~~~~~~~~~~~-~ 4  \phi_{\a \b} [\psi_a {}^{\a j} \Bar
\psi_b{}^\b{}_i ~-~ \frac 12 \d_i^j \psi_a {}^{\a k} \Bar
\psi_b{}^\b{}_k ] \\             
& & ~~~~~~~~~~~~~~~~~~-~ 2 (\g^3)_{\a \b} [~\Bar B ( C_{i k} \psi_a {}^{\a k} 
\psi_b {}^{\b}{}^{k} ~-~ \frac 12 \d_i^j C_{k l} \psi_a {}^{\a k} \psi_b 
{}^{\b l})
\\ & & ~~~~~~~~~~~~~~~~~~+~ B ( C^{j k} \Bar \psi_a {}^{\a}{}_{i}
 \Bar \psi_{b}{}^{ \b}{}_{k} ~-~ \frac 12 \d_i^j  C^{k l} \Bar \psi_a{}^{\a}
{}_{k} \Bar \psi_{b}{}^{\b}{}_{l})~ ] ~{\large \} } ~~ ,   \end{array}      
\label{FSs}  
\end{equation}
where $\e^{a b} \, {\car}_{a b} (\hat \omega)$ is the usual two-dimensional curvature in
terms of ${\rm e}_a {}^m$ and $\hat \omega_m $. 

\section{2D, ${\cal N}=(2,2)$ Local Superspace
 Integration \& Chiral Projector}

~~~~ In a rigid superspace (i.e. in a ``flat supergravity background''), the derivation of
component results follows most efficiently \cite{ggrs1} from replacing the integration
of fermionic coordinates by a process using {\em {first}} application of the superspace 
covariant derivative followed by taking a limit in which all Grassmann coordinates are
taken to zero.  For the discussion of this paper, this amounts to the validity of the following
equation,
\be \eqalign{
 S \ &= \  \int d^2 \s\ d^4 \theta\ \,  d^4 {\Bar \theta\ } \,  {\cal L} ~=~ \lim_{\q \to 0}
\int d^4 \s ~ \, \fracm 12 \,  {\Big [} \,  {\rm {D}}{}^4 \,  {\rm {\Bar D}}{}^4 \,  \,  {\cal L} ~+~ 
{\rm h}. \, {\rm c}. \, {\Big ]}   \cr
~&\equiv~ 
\int d^2 \s ~ \, \fracm 12 \,  {\Big [} \,  {\rm {D}}{}^4 \,  {\rm {\Bar D}}{}^4 \,  \,  {\cal L} ~+~ 
{\rm h}. \, {\rm c}. \, {\Big ]} \, {\Big |}
  } \label{Sint} \ee

In the presence of a supergravity background the integration over the superspace 
measure is modified by the insertion of the supergravity vielbein
\be \eqalign{
\int d^2 \s\ d^4 \theta\ \,  d^4 {\Bar \theta\ } ~ &\to ~  \int d^2 \s\ d^4 \theta\ \, 
d^4 {\Bar \theta\ } \,  {\rm E}{}^{-1}
  } \label{Sint2} \ee
and this in turn implies a modification of (\ref{Sint2}) to the form \cite{Ecto}
\be \eqalign{
S ~&=~ \int d^2 \s ~ \, \fracm 12 \,  {\rm e}{}^{-1}  \, {\Big [} \,  
  {\cal {D}}{}^{(4{\cal P})}  \,  {\cal {\Bar D}}{}^{(4{\chi})}  \,
  \,  {\cal L} ~+~ {\rm h}. \, {\rm c}. \, {\Big ]} \, {\Big |}
  }    \label{Sint3} \ee 
which is written in terms of two differential operators, ${\cal {\Bar D}}^{(4)}_{{\cal P}}$ 
and $ {\cal {\Bar D}}_{{\chi}}^{(4)}$.  The first of these is called the ``density projection 
operator'' and the second is called ``chiral projection operator.''   The appearance of these
two distinct operators is characteristic of any superspace in which chirality has a well defined
meaning.

As outlined in  \cite{Ecto}, the density projection operator must be of the
form 
\be
{\cal {D}}{}^{(4{\cal P})} ~=~ \sum_{i = 0}^{4} \, b_{( 4 - i)} \,  \cdot\,  \, \left[ \,
(\nabla) \, \times\,  \, \cdots\, \, \times\,( \nabla)^{4 - i}  \, \right]
~~~,  \label{exp1} 
 \ee
in terms of some field-dependent coefficients $b_{(4 - i)}$ and powers of the spinorial superspace 
supergravity covariant derivative $\nabla_{\a \, i}$.  In the work of \cite{ggrs1}, a `handicraft' method
for finding these coefficients was described.   The works of \cite{Ecto} describe more powerful
methods for deriving this operator (which will be a topic of future efforts).   In a similar manner, 
the chiral projection operator must be of the form 
\be
 {\cal {\Bar D}}{}^{(4{\chi})} ~=~ \sum_{i = 0}^{4} \, a_{( 4 - i)} \, 
\cdot\,  \, \left[ \, (\Bar \nabla) \, \times \, \cdots\,  \, \times
\,({\Bar  \nabla})^{4 - i}  \, \right]
~~~,
 \label{exp2} 
 \ee
in terms of some field-dependent coefficients $a_{(4 - i)}$ and powers of the spinorial superspace 
supergravity covariant derivative ${\Bar \nabla}{}_{\a}^{ \, i}$. 

Here we give our main result for the local 2D, $\cal N$ = (2, 2) superspace described by (\ref{CommAlg})
\be \eqalign{
 S \ =&\  \int d^2 \s\ d^4 \theta\ \,  d^4 {\Bar \theta\ } \,  {\rm E}{}^{-1} \, {\cal L} \cr
  \ =&\  \int d^2 \s\ d^4 \theta\ \,  \,  {\cal E}^{-1} \, \fracm 12 \, 
   \left[ \, {\Bar {\nabla}}{}^{(2) \, \a \, \b} 
 ~-~ 2 \, B \, (\g^3){}^{\a \, \b} \, \right] \, {\Bar {\nabla}}{}^{(2)}_{ \, \a \, \b} 
  \, {\cal L} \, {\Big |} ~+~ {\rm h}. \, {\rm c}. \cr
  \ =&\  \int d^2 \s\  \, \,   {\rm e}^{-1} \,  \fracm 12 \,  {\cal {D}}_{{\cal P}}^{(4)} \, 
   \left[ \, {\Bar {\nabla}}{}^{(2) \, \a \, \b} 
 ~-~ 2 \, B \, (\g^3){}^{\a \, \b} \, \right] \, {\Bar {\nabla}}{}^{(2)}_{ \, \a \, \b} 
  \, {\cal L} \, {\Big |} ~+~ {\rm h}. \, {\rm c}. \cr
  } \label{Dense} \ee
where on the first line $ {\rm E}^{-1} $ is the density factor (i.e. the superdeterminant of the 
vielbein of the full 2D, $\cal N$ = (2, 2) superspace) and on the second line of this equation
$ {\cal E}^{-1} $ is the chiral density factor.  The superdifferential operator
\be \eqalign{
 {\cal {\Bar D}}{}^{(4{\chi})} ~=~ 
 \left[ \, {\Bar {\nabla}}{}^{(2) \, \a \, \b} 
 ~-~ 2 \, B \, (\g^3){}^{\a \, \b} \, \right] \, {\Bar {\nabla}}{}^{(2)}_{ \, \a \, \b} 
 } \label{ChRLprj}  \ee
is the 2D, $\cal N$ = (2, 2) chiral projection operator.  The main new result we have to
report is its explicit form which satisfies
\be \eqalign{ {\Bar \nabla}{}_{\g}^{ i} \,   {\cal {\Bar D}}{}^{(4{\chi})}   \, \Psi 
~=~ {\Bar \nabla}{}_{\g}^{ i} \,  \left[ \, {\Bar {\nabla}}{}^{(2) \, \a \, \b}  ~-~ 2 \, B \, 
(\g^3){}^{\a \, \b} \, \right] \, {\Bar {\nabla}}{}^{(2)}_{ \, \a \, \b} \, \Psi  ~=~ 0
 } \label{ChRLcond}   \ee
for any general scalar superfield $ \Psi$.  

Upon comparing (\ref{exp2}) with (\ref{ChRLprj}), we note the coefficients for the former can 
be read from the latter and imply that 
\be \eqalign{
 {~~~~~~~~~}    
 a_{(0)} ~&=~ 0  ~~~~,  ~~~~
 a_{(1)} ~=~ 0  ~~~~,  ~~~~
 a_{(2)} ~=~ - \, 2 \, B \, (\g^3){}^{\a \, \b} \, C{}_{i \, j}  ~~~~,  \cr
 a_{(3)} ~&=~ 0  ~~~~,  ~~~~
 a_{(4)} ~=~  \fracm 12 \, C{}_{i \, j} \,  C{}_{k \, l} \, \left[ \, C{}^{\a \, \g} \, C{}^{\b \, \d} ~+~
 C{}^{\a \, \d} \, C{}^{\b \, \g}  \, \right]   ~~~~.  \cr
 } \label{a-coeff}   \ee
So that explicitly we have
 \be \eqalign{
 {\cal {\Bar D}}{}^{(4{\chi})} ~=~  &\fracm 12 \, C{}_{i \, j} \,  C{}_{k \, l} \, 
 \left[ \, C{}^{\a \, \g} \, C{}^{\b \, \d} ~+~ C{}^{\a \, \d} \, C{}^{\b \, \g}  \, \right]  
 {\Bar \nabla}{}_{\a}^i \, {\Bar \nabla}{}_{\b}^j \, {\Bar \nabla}{}_{\g}^k \,
 {\Bar \nabla}{}_{\d}^l  {~~~~~~~}   {~~~~~}  \cr
 & - \, 2 \, B \, (\g^3){}^{\a \, \b} \, C{}_{i \, j} \,  {\Bar \nabla}{}_{\a}^i \, {\Bar \nabla}{}_{\b}^j
~~~.
} \label{exp3} 
 \ee
Even without knowing the explicit form of the density projection operator, (\ref{Dense}) implies
\be \eqalign{
  \int d^2 \s\ d^4 \theta\ \,  d^4 {\Bar \theta\ } \,  {\rm E}{}^{-1} ~=~ 0 ~~~,
  } \label{Vol=0} \ee
i.e. the superspace described by (\ref {CommAlg}) has a vanishing supervolume. The derivation
of these results are described in an appendix.

\section{Conclusion }

~~~~ With this present work, we have completed half of the task of developing an efficient
local superspace integration theory for two dimensional theories that possess eight supercharges.
The crux of this presentation was the unveiling of the explicit form of the chiral projection
operator given in (11).

 \vspace{.1in}
 \begin{center}
 \parbox{4in}{{\it ``Everone takes the limits of his own vision for the limits $\,~~$ of the world.''}\,\,-\,\,Arthur Schopenhauer}
 \end{center}

 \vspace{.2in}

 \noindent

 {\bf Acknowledgements}\\[.1in] \indent
This research was supported in part by the endowment of the John S.~Toll Professorship,
the University of Maryland Center for String \& Particle Theory, National Science Foundation 
Grant PHY-0354401.  AM wishes to acknowledge the hospitality of the University of Maryland 
and in particular of the Center for String and Particle Theory. Additional he wishes to recognize 
his participation in the 2008 SSTPRS (Student Summer Theoretical Physics Research Session).

\newpage
\noindent
{\Large{\bf Appendix A:  Definitions \& Conventions}}

~~~~ For two dimensional superspaces, we use the following conventions for the
quantities associated with spinors.

$$ \eta_{ a b} = (1 , -1)  ~~~,~~~ \e_{a b} \e^{c d} ~=~  - \d_{[a}{}^c
  \d_{b]}{}^d  ,~~~ \e^{0 1} =  +1 ~~~, $$
$$  (\g^a)_{\a}{}^{ \g} (\g^b)_{\g}{}^{ \b} =   \eta^{ a b} \d_{\a}{}^{\b}
- \e^{ a b}  (\g^3)_{\a}{}^{\b} ~~.  \eqno(A.1)  $$
The last one of these relations imply
$$ \eqalign{
\g^a \g_{a} &=~ 2 \, {\bf I} ~~~, ~~~
\g^3 \g^{a} ~=~ - \e^{a \, b} \g_{b} ~~~. }  \eqno(A.2) 
$$

Some useful Fierz identities are:
$$ C_{\a \b}  C^{\g \d} ~=~  \d_{[\a}{}^\g  \d_{\b]}{}^\d ~~, $$
$$ (\g^a)_{\a \b} (\g_a)^{\g \d} ~+~ (\g^3)_{\a \b} (\g^3)^{\g \d}
~=~ - \d_{(\a}{}^\g  \d_{\b)}{}^\d  ~~, $$
$$ (\g^a)_{(\a}{}^\g (\g_a)_{\b)}{}^\d ~+~ (\g^3)_{(\a}{}^\g 
(\g^3)_{\b)}{}^\d ~=~ \d_{(\a}{}^\g  \d_{\b)}{}^\d    ~~ ,  $$
$$ (\g^a)_{(\a}{}^\g (\g_a)_{\b)}{}^\d ~=~ -2 (\g^3)_{\a \b} 
(\g^3)^{\g \d} ~~, $$
$$ 2 (\g^a)_{\a \b} (\g_a)^{\g \d} ~+~ (\g^3)_{(\a}{}^\g (\g^3)_{\b)}
{}^\d   ~=~ - \d_{(\a}{}^\g  \d_{\b)}{}^\d ~~,$$
$$ (\g_a)_\a {}^\d \d_\b {}^\g ~+~ (\g^3 \g_a)_\a {}^\g (\g^3)_\b {}^\d
~=~ (\g^3 \g_a)_{\a \b} (\g^3)^{\g \d} ~~.    \eqno(A.3) $$

In terms of an explicit representation, we can define the 2D $\g$-matrices
in terms of the usual Pauli matrices according to
$$
(\g^0 )_{\a}{}^{\b} ~\equiv~ (\s^2  )_{\a}{}^{\b}  ~~~,~~~ (\g^1 
 )_{\a}{}^{\b} ~\equiv~ - i (\s^1  )_{\a}{}^{\b} ~~~,~~~
\g^3 ~\equiv~ (\s^3  )_{\a}{}^{\b} ~~~.  \eqno(A.4) $$
As can be seen, these satisfy the second line in (A.1). The spinor metric
$C_{\a \b}$ and its inverse $C^{\a \b}$ can be identified as
$$
C_{\a \b} ~\equiv~ (\s^2 )_{\a \, \b} ~~~,~~~ C^{\a \b} ~\equiv~ 
- (\s^2 )^{\a \, \b} ~~~. \eqno(A.5) 
$$
Using this explicit representation, it is easy to show the following
symmetry properties
$$ \eqalign{ {~~~~~~~}
(\g^a)_{\a \b} ~=~ (\g^a)_{\b \a} ~~~,~~~
(\g^3)_{\a \b} ~=~ (\g^3)_{\b \a} ~~~, ~~~
C_{\a \b} ~ ~=~ - C_{\b \a} ~~~, \cr
(\g^a)^{\a \b} ~=~ (\g^a)^{\b \a} ~~~,~~~
(\g^3)^{\a \b} ~=~ (\g^3)^{\b \a} ~~~, ~~~
C^{\a \b} ~ ~=~ - C^{\b \a} ~~~. 
} \eqno(A.6)    $$
In a similar manner the following complex conjugation properties
can be derived
$$
[(\g^a)_{\a}{}^{ \b} ]^* ~=~ - \, (\g^a)_{\a}{}^{ \b} ~~~,~~~
[(\g^3)_{\a}{}^{ \b} ]^* ~=~ + \, (\g^3)_{\a}{}^{ \b} ~~~,
 \eqno(A.7) 
$$
$$ \eqalign{ {~~~~~~~}
[ (\g^a)_{\a \b}  ]^* ~=~ (\g^a)_{\a \b} ~~~,~~~
[ (\g^3)_{\a \b}  ]^* ~=~ -\, (\g^3)_{\a \b} ~~~, ~~~
[ C_{\a \b}  ]^* ~=~ - C_{\a \b} ~~~, \cr
[ (\g^a)^{\a \b}  ]^* ~=~ (\g^a)^{\a \b} ~~~,~~~
[ (\g^3)^{\a \b}  ]^* ~=~ -\, (\g^3)^{\a \b} ~~~, ~~~
[ C^{\a \b}  ]^* ~=~ - C^{\a \b} ~~~. 
} \eqno(A.8)    $$

Due to the first relation in (A.8) we see that this choice of gamma matrices is
in a Majorana representation and thus the simplest spinors such as $\psi^{\a}(x)$, 
may be chosen to be real, i.e.
$$
[\, \psi^{\a}(x) \, \, ]^* ~=~ \psi^{\a}(x) ~~~,
 \eqno(A.9)  
$$ 
and we can raise and lower spinor indices according to
$$
 \psi^{\a}(x) ~=~ C{}^{\a \, \b} \,  \psi_{\b}(x)  ~~~,~~~
  \psi_{\a}(x) ~=~  \psi^{\b}(x)  \,  C{}_{\b \, \a}  ~~~.
 \eqno(A.10)  $$
It can be seen as a consequence that
$$
[\, \psi_{\a}(x) \, \, ]^* ~=~   - \, \psi_{\a}(x)  ~~~.
 \eqno(A.11)  
$$
Of course, it is always possible to introduce complex spinors also.

The two generators that define the holonomy group of the 2D, $\cal N$ = (2, 2)
superspace are $\cal M$ and ${\cal Y}_i{}^j$, respectively for the 2D Lorentz group
(SL(2,R)) and an internal SU(2) group.  These are defined to act as
$$
{\Big [} \, {\cal M} ~,~ \nabla_{\a \, i} \, {\Big]} ~=~ \fracm 12 \, 
(\g^3)_{\a} {}^{\b} \, \nabla_{\b \, i}    ~~~,~~~
{\Big [} \, {\cal Y}{}_i{}^j ~,~ \nabla_{\a \, k} \, {\Big]} ~=~ \d{}_k{}^j \, \nabla_{\a \, i} 
~-~ \fracm 12 \,  \d{}_i{}^j \,  \nabla_{\a \, k} 
 \eqno(A.12) 
$$

The rules for maniplulating the SU(2) spinors are very much similar to the ones
used for the SL(2,R) spinor indices.  The SU(2) metric
$C_{i \, j}$ and its inverse $C^{i \, j}$ can be identified as
$$
C_{i \, j} ~\equiv~ (\s^2 )_{i \, j} ~~~,~~~ C^{i \, j} ~\equiv~ 
- (\s^2 )^{i \, j} ~~~,  \eqno(A.13) 
$$
so that
$$
C_{i \, j} ~ = ~ - \, C_{j \, i} ~~,~~ C^{i \, j} ~ = ~ - \, C^{j \, i} ~~,~~
~~ C_{i \, j}  \, C^{k \, l} ~=~ \d{}_i{}^k \,  \d{}_j{}^l ~-~  \d{}_i{}^l \,  \d{}_j{}^k
 ~~~. \eqno(A.14) 
$$
We raise and lower SU(2)  indices according to
$$
 \psi^{i}(x) ~=~ C{}^{i \, j} \,  \psi_{j}(x)  ~~~,~~~
  \psi_{\i}(x) ~=~  \psi^{j}(x)  \,  C{}_{j \, i}  ~~~,
 \eqno(A.15)  $$
that are directly the analogs for raising and lowering indices on SL(2,R) tensors.

\newpage
{\Large{\bf Appendix B:  Derivation Of Chiral Projector}}

~~~~ In this appendix, we give a presentation that leads to the form of
the chiral projector.

The expressions in (\ref{CommAlg}) give the commutator algebra of the 4D, $\cal N$
$=$ (2, 2) supergravity derivative.  However, in order to derive the chiral projection
formula we need to go beyond that algebra.  For this purpose, we introduce new
second order differential operators denoted by
$\nabla{}^{(2)}_{\a \, \b}   ~\& ~  \nabla{}^{(2)}_{i \, j} 
$
and defined by the equations
 $$
 \eqalign{ 
~~&\nabla{}^{(2)}_{\a \, \b}  ~=~  \fracm 12 \, C^{i \, j} \left[ \,
\nabla_{\a \, i} \, \nabla_{\b \, j}  ~+~ \nabla_{\b \, i} \, \nabla_{\a \, j} 
\, \right] ~~~, \cr 
~~&  \nabla{}^{(2)}_{i \, j}   ~=~  \fracm 12 \, C^{\a \, \b} \left[ \,
\nabla_{\a \, i} \, \nabla_{\b \, j}  ~+~ \nabla_{\a \, j} \, \nabla_{\b \, i} 
\, \right] ~~~.
}  \eqno(B.1)  $$
Using these definitions implies
 $$
 \eqalign{  {~~}
 \nabla_{\a \, i} \, \nabla_{\b \, j}  ~&=~   \fracm 12  \left[ \,  \nabla_{\a \, i} ~,~
  \nabla_{\b \, j}  \, \right]   ~+~  \fracm 12 \left\{ \,  \nabla_{\a \, i} ~,~
  \nabla_{\b \, j}  \, \right\}   ~~~, \cr
  ~&=~   \fracm 12 \, C_{i \, j} \, \nabla{}^{(2)}_{\a \, \b} ~+~  \fracm 12 \, C_{\a \, \b} \, 
  \nabla{}^{(2)}_{i \, j} ~+~  \fracm 12 \left\{ \,  \nabla_{\a \, i} ~,~
  \nabla_{\b \, j}  \, \right\}   ~~~, \cr
  ~&=~   \fracm 12 \, C_{i \, j} \, \nabla{}^{(2)}_{\a \, \b} ~+~  \fracm 12 \, C_{\a \, \b} \, 
  \nabla{}^{(2)}_{i \, j} ~+~ \Bar B  \, [ \, C \sb {\a \b} C \sb{ ij} \cm ~-~ 
   ( \g \sp 3 ) \sb{ \a \b} \cy \sb{i j} \, ] 
 ~~~.
}     \eqno(B.2)  $$
Next we use (\ref{CommAlg}), (B.1) and (B.2) to derive 
$$
 \eqalign{  {~~~}
\left[ \, \nabla{}_{\a \, i} ~,~  \nabla{}^{(2)}_{\b \, \g} \, \right] ~=~ &- \, \fracm 14 \, 
 \Bar B  \, [ \, C \sb {\a \, ( \b}  ( \g \sp 3 ) \sb{ \g )}{}^{\d} ~-~ 3 \, 
    ( \g \sp 3 ) \sb{ \a ( \b} \,  \d{}_{\g )}{}^{\d} \, ]  \, \nabla{}_{\d \, i}  \cr
 &- \,  \, \Bar B  \, \nabla{}_{(\b \, j} \, 
 [ \, C \sb {\a \, | \g)}  \d{}_i{}^j \cm ~-~    ( \g \sp 3 ) \sb{ \a \, | \g )} \cy {}_i{}^j \, ] 
~~~. }  
\eqno(B.3) $$
Using more manipulations there are found additional
identities involving $\nabla{}^{(2)}_{\a \, \b} $. 
$$ 
 \nabla{}_{( \a | \, i}  \nabla{}^{(2)}_{| \b \, \g ) }  ~=~ 2\, 
\Bar B  \, ( \g \sp 3 ) {}_{( \a \, \b}  \,  \left[ \, \nabla{}_{\g ) \, i} ~+~ 
 \nabla{}_{\g ) \, j} \cy {}_i{}^j \, \right] 
~~~,  {~~~~~~~~~}   {~~~~~~~~~}    {~~~~~~~~}
 \eqno(B.4) $$
$$ 
 \eqalign{  {~~~\,~}
 \nabla{}^{(2)}_{\a \, \b} \,  \nabla{}^{(2) \, \b \, \g}  ~=~ \fracm 12 \,
  \d{}_{\a}{}^{\g} \, 
\nabla{}^{(2) \,  \b \, \d} \,  \nabla{}^{(2)}_{ \b \, \d}  ~+~  \fracm 12 \,
C{}^{\b \, \d} \, C{}^{\g \, \e}
\left[ \, \nabla{}^{(2)}_{\a \, \b} ~,~  \nabla{}^{(2)}_{\d \, \e} \, \right] 
~~~, {~~}}  
\eqno(B.5) $$
$$
 \eqalign{ 
\left[ \, \nabla{}^{(2)}_{\a \, \b} ~,~  \nabla{}^{(2)}_{\d \, \e} \, \right] ~&=~ 2\,
 \Bar B  \, \left[ \,  ( \g \sp 3 ) \sb{ \a \, \b } \,   \nabla{}^{(2)}_{\d \, \e} ~-~
  ( \g \sp 3 ) \sb{ \d \, \e } \,   \nabla{}^{(2)}_{\a \, \b}  \, \right]  \cr
 &~~~~+\, 4 \, {\Bar B}^2  \, \left[ \, C \sb {\a \, ( \d}  ( \g \sp 3 ) \sb{ \e )  \, \b}  
 ~+~  C \sb {\b \, ( \d}  ( \g \sp 3 ) \sb{ \e )  \, \a}{} \,  \right] \cm   \cr
 &~~~~-\,  {\Bar B}  \, \left[ \, C \sb {\a \, ( \d}  \nabla^{(2)} \sb{ \e )  \, \b}  
 ~+~  C \sb {\b \, ( \d}  \nabla^{(2)}  \sb{ \e )  \, \a}{} \,  \right] \cm   \cr
 &~~~~+\, \fracm 12 \, {\Bar B}  \,  \nabla^{(2)}{}_{i \, j}
  \left[ \, C \sb {\a \, ( \d}  ( \g \sp 3 ) \sb{ \e )  \, \b}  
 ~+~  C \sb {\b \, ( \d}  ( \g \sp 3 ) \sb{ \e )  \, \a}{} \,  \right] \cy {\,}^{i \, j}
 ~~~. 
 {~~~~~~~~}  {~~~~~~~~~~}    }  
\eqno(B.6) $$ 
With the identities of (B.4) - (B.6) in hand, the proof of (11) follows from the steps 
described below.
\newline $~$ \newline 
\noindent
(1.) The first step is to multiply (B.4) from the right by $ \nabla{}^{(2)}{}^{ \b \, \g}{\Bar \Psi}$
which yields 
$$  \eqalign{ {~~~~~~}
 &\nabla{}_{ \a  \, i}  \nabla{}^{(2)}_{ \b \, \g  } \,  \nabla{}^{(2)}{}^{ \b \, \g}{\Bar \Psi}
 ~+~ 2 \,  \nabla{}_{ \g  \, i}  \nabla{}^{(2)}_{ \a \, \b  } \,  \nabla{}^{(2)}{}^{ \b \, \g}{\Bar \Psi}
 ~-~  2\, \Bar B  \, ( \g \sp 3 ) {}_{ \b \, \g} \, \nabla{}_{\a \, i} \,  \nabla{}^{(2)}{}^{ \b \, \g}{\Bar 
 \Psi} \cr
 &~-~  4\, \Bar B  \, ( \g \sp 3 ) {}_{ \a \, \b} \, \nabla{}_{\g \, i} \,  \nabla{}^{(2)}{}^{ \b \, \g}{\Bar 
 \Psi} ~=~  0 ~~~. 
} \eqno(B.7)
$$
\newline $~$ \newline 
\noindent
(2.) In the second term of (B.7), we substitute the identity from (B.5) to find
$$  \eqalign{ {~~~~~~~~~}
 &2 \nabla{}_{ \a  \, i}  \nabla{}^{(2)}_{ \b \, \g  } \,  \nabla{}^{(2)}{}^{ \b \, \g}{\Bar \Psi}
 ~+~  C{}^{\b \, \d} \, C{}^{\g \, \e}  \,  \nabla{}_{ \g  \, i} \,
\left[ \, \nabla{}^{(2)}_{\a \, \b} ~,~  \nabla{}^{(2)}_{\d \, \e} \, \right]  \, {\Bar \Psi}
 \cr
 &~-~  2\, \Bar B  \, ( \g \sp 3 ) {}_{ \b \, \g} \, \nabla{}_{\a \, i} \,  \nabla{}^{(2)}{}^{ \b \, 
 \g}{\Bar \Psi}  ~-~  4\, \Bar B  \, ( \g \sp 3 ) {}_{ \a \, \b} \, \nabla{}_{\g \, i} \,  \nabla{}^{
 (2)}{}^{ \b \, \g}{\Bar  \Psi} ~=~  0 ~~~. 
} \eqno(B.8)
$$
\newline $~$ \newline 
\noindent
(3.) We next multiply the identity of (B.5) from the right by $\Bar \Psi$ to find
$$
 \eqalign{ 
\left[ \, \nabla{}^{(2)}_{\a \, \b} ~,~  \nabla{}^{(2)}_{\d \, \e} \, \right] \, {\Bar \Psi}
~=~ 2\,
 \Bar B  \, \left[ \,  ( \g \sp 3 ) \sb{ \a \, \b } \,   \nabla{}^{(2)}_{\d \, \e} ~-~
  ( \g \sp 3 ) \sb{ \d \, \e } \,   \nabla{}^{(2)}_{\a \, \b}  \, \right] \, {\Bar \Psi} ~~~.
 }  
 \eqno(B.9)
$$
\newline $~$ \newline 
\noindent
(4.) The substitution of the result on the left hand side of (B.9) into the second term
 $~~~~~~$ of (B.8) yields
$$  \eqalign{ {~~~~~~~~~~}
 &2 \nabla{}_{ \a  \, i}  \nabla{}^{(2)}_{ \b \, \g  } \,  \nabla{}^{(2)}{}^{ \b \, \g}{\Bar \Psi}
 ~-~  2\, \Bar B  \, ( \g \sp 3 ) {}^{ \b \, \g} \, \nabla{}_{\g \, i} \,  \nabla{}^{
 (2)}{}_{ \a \, \b}{\Bar  \Psi} \cr
 &~-~  2\, \Bar B  \, ( \g \sp 3 ) {}_{ \b \, \g} \, \nabla{}_{\a \, i} \,  \nabla{}^{(2)}{}^{ \b \, 
 \g}{\Bar \Psi}  ~-~  2\, \Bar B  \, ( \g \sp 3 ) {}_{ \a \, \b} \, \nabla{}_{\g \, i} \,  \nabla{}^{
 (2)}{}^{ \b \, \g}{\Bar  \Psi} ~=~  0  ~~~,  \cr
  &{~~}
   \cr
  &2 \nabla{}_{ \a  \, i}  \nabla{}^{(2)}_{\b \, \g  } \,  \nabla{}^{(2)}{}^{ \b \, \g}{\Bar \Psi}
 ~+~  2\, \Bar B  \, ( \g \sp 3 ) {}^{\g}{}_{\b} \, \nabla{}_{\g \, i} \,  \nabla{}^{
 (2)}{}_{ \a}{}^{\b}{\Bar  \Psi} \cr
 &~-~  2\, \Bar B  \, ( \g \sp 3 ) {}_{ \b \, \g} \, \nabla{}_{\a \, i} \,  \nabla{}^{(2)}{}^{ \b \, 
 \g}{\Bar \Psi}  ~-~  2\, \Bar B  \, ( \g \sp 3 ) {}_{ \a \, \b} \, \nabla{}_{\g \, i} \,  \nabla{}^{
 (2)}{}^{ \g \, \b}{\Bar  \Psi} ~=~  0  ~~~.
} \eqno(B.10)
$$
\newline $~$ \newline 
\noindent
(5.) The second and last terms on the final line in (B.10) added together equal to the
\newline $~~~~~~$  third term in the same equation. Thus we obtain the final result
$$  \eqalign{ {~~~~~~~~~~~~~}
&2 \nabla{}_{ \a  \, i}  \nabla{}^{(2)}_{\b \, \g  } \,  \nabla{}^{(2)}{}^{ \b \, \g}{\Bar \Psi}
 ~-~  4\, \Bar B  \, ( \g \sp 3 ) {}_{ \b \, \g} \, \nabla{}_{\a \, i} \,  \nabla{}^{(2)}{}^{ \b \, 
 \g}{\Bar \Psi}  ~=~  0  \cr
 &2 \nabla{}_{ \a  \, i}  \left( \, \nabla{}^{(2)}_{\b \, \g  } \,  \nabla{}^{(2)}{}^{ \b \, \g}
 ~-~  2\, \Bar B  \, ( \g \sp 3 ) {}_{ \b \, \g} \,  \nabla{}^{(2)}{}^{ \b \, 
 \g} \, \right) \, {\Bar \Psi}  ~=~  0  \cr
 & \nabla{}_{ \a  \, i}  \left( \, \nabla{}^{(2)}{}^{\b \, \g  } 
 ~-~  2\, \Bar B  \, ( \g \sp 3 ) {}^{ \b \, \g} \, \right) \,  \nabla{}^{(2)}{}_{ \b \, 
 \g} \, {\Bar \Psi}  ~=~  0 
 ~~~.
} \eqno(B.11)
$$
 Upon taking the complex conjugate of this final equation in (B.11) the result of (11) follows.

\end{document}